\documentstyle[aps,twocolumn]{revtex}
\input epsf 

\begin{document}

\twocolumn[\hsize\textwidth\columnwidth\hsize\csname@twocolumnfalse%
\endcsname
\title{Fractional vortices on grain boundaries --- the case for broken
time reversal symmetry in high temperature superconductors}

\author{D.B. Bailey}
\address{Department of Physics, Stanford University, Stanford, CA 94305}

\author{M. Sigrist}
\address{Department of Physics, Massachussets Institute of Technology,
        Cambridge, MA 02139 \\
        and \\
        Theoretische Physik, ETH-H\"onggerberg, 8093 Z\"urich, Switzerland}

\author{and \\ R. B. Laughlin}
\address{Department of Physics, Stanford University, Stanford, CA 94305\\
        and \\
        Lawrence Livermore National Laboratory,
        P. O. Box 808, Livermore, CA 94550}

\maketitle

\begin{abstract}
We discuss the problem of broken time reversal symmetry near grain boundaries
in a $d$-wave superconductor based on a Ginzburg-Landau theory. It is
shown that such a state can lead to fractional vortices on the grain
boundary.  Both analytical and numerical results show the structure of
this type of state.
\end{abstract}

]

\vspace{0.3in}

\baselineskip=0.385 cm

\section{Introduction}

During the last few years the understanding of the
microscopic properties of the high-temperature
superconductors (HTSC) has gradually improved
\cite{LOSALAMOS}.
While for a long time studies have focused on the
exotic normal state properties, recently the
interest turned more towards the superconducting
phase, in particular, the symmetry of the order
parameter. For a tetragonal system the list of
possible order parameter symmetries is rather
long \cite{SYMMETRY}. However, the recent debate has essentially
concentrated only on two symmetries of the Cooper
pair wavefunction \cite{LEVI}. One is due to ``$s$-wave'' pairing,
the most symmetric pairing channel. The other
is ``$d$-wave'' pairing, where the pair wavefunction
($ \psi({\bf k}) \propto \cos k_x - \cos k_y $)
changes sign under $ 90^o $-rotations in the basal
plane of the tetragonal crystal lattice. As a consequence,
the latter wave function has nodes along the
[110]-direction. A possible alternative to
the standard $s$-wave was presented with the ``extended
$s$-wave'' pairing state ($ \psi({\bf k}) \propto
\cos k_x + \cos k_y $) which has also nodes in the
first Brillouin zone, but is completely symmetric under
all operations of the tetragonal point group $ D_{4h} $
\cite{EXTENDEDS}.
For the orthorhombically distorted system, the $s$- and
$d$-wave channels are not distinguished by symmetry.
Nevertheless, we expect that basic properties of
the pair wave function such as the existence of sign changes
and nodes are retained if they were present in the
tetragonal case \cite{ABRIKOSOV}.

A variety of experiments have been performed in order to
distinguish among the order parameter symmetries. One class
of experiments considers the properties of the quasi
particle excitations in the superconducting state. The
existence of nodes in the pair wavefunction implies that
there are also nodes in the excitation gap.
Low-lying excitations at the nodes modify the low-temperature
behavior of certain thermodynamic properties compared with
that of a superconductor which opens a complete gap.
The clearest sign of such an effect was observed for the
London penetration depth which behaves as $ \lambda(T)
-\lambda(0) \propto T $ in contrast to the conventional
exponential law, $ \propto \exp(- \Delta/k_B T) $
\cite{HARDY}. This result
strongly suggests that there are nodes in the gap and the
pair wavefunctions, which are compatible with both extended
$s$-wave and $d$-wave as well as with a very anisotropic $s$-wave
state.

Another class of experiments aims at the direct observation of the
intrinsic phase structure, the sign changes of the pair
wavefunction. The Josephson effect as a phase coherent coupling
of the order parameters of two superconductors provides the
natural means for this purpose \cite{GESH,SRWOHL}. Arrangements connecting
YBCO single crystals at two perpendicular surfaces to a
standard $s$-wave superconductor to form a loop for a
SQUID have been
used to detect a phase difference between the $ x $- and
$ y $-direction of the pair wavefunction \cite{SQUID}. The experiments
observe with good precision a phase difference of $ \pi $
compatible with the $ d_{x^2-y^2} $-wave order parameter.

The intrinsic $\pi$ phase shift in this configuration
leads to frustration effects which manifest themselves
in the form of a spontaneous supercurrent flowing
around the loop.  The supercurrent generates a flux $ \Phi = \pm \Phi_0/2 $
where $ \Phi_0 = hc/2e $ is the standard flux quantum. This property
has recently been detected and the flux was measured
with very high accuracy \cite{RING}.

On the other hand, several other experiments based on the
Josephson effect seem at present to contradict the presence of a $d$-wave
order parameter. Chaudhari and Lin analyzed the Josephson
current through a grain boundary in the basal plane
with a special geometry giving a
basal plane contact between two segments of a YBCO film
\cite{CHAUD}.
They demonstrated that various properties might support
an order parameter with $s$-wave rather than simple $d$-wave symmetry.
The interpretation of this experiment, however, has recently
been contested by Millis \cite{MILLIS}. In contrast, Sun and coworkers
investigated
Josephson tunneling between a standard $s$-wave superconductor (Pb)
and YBCO, where the tunneling direction is the $c$-axis of YBCO
\cite{DYNES}.
These data so far could not be explained consistently within
the picture of pure $d$-wave superconductivity. Therefore,
the simple $d$-wave scenario may not be sufficient for a complete
understanding of all experiments introduced here \cite{SKLMR}.

Indeed a recent experiment by Kirtley and coworkers suggests that the
situation is more complicated than might be naively expected for
a $d$-wave superconductor \cite{TRIANGLE}.
Their experimental arrangement consists
of two segments of $c$-axis textured YBCO films where one
is a triangular inclusion within in the other. The basal plane
crystalline axes are misaligned with one another.
The boundary of the triangle acts as junction between the two
segments. We will show in Section II that if YBCO were a
$d$-wave superconductor
we would expect vortices to appear spontaneously at two of the three
corners of the triangle each containing a flux of $ \pm \Phi_0 /2 $.
The experiment does find spontaneous vortices at corners, but
these vortices have fluxes different from $ n\Phi_0/2 $
($ n $: integer). In addition, flux appears at all three corners
and occasionally also on an edge of the triangle. We will argue in
Section II that this can be explained by a superconducting state
which violates time-reversal symmetry $ {\cal T} $.
Therefore, the simple picture of a single component
$ d_{x^2-y^2} $-wave order parameter might not apply here.

$ {\cal T} $-violation is not uncommon in the field of unconventional
superconductivity. A large number of superconducting states classified
by symmetry indeed break time-reversal symmetry
\cite{SYMMETRY,REVIEW}. In the complex
superconducting phase diagrams of the heavy fermion
compounds, $ {\rm UPt}_3 $ and $ U_{1-x} {\rm Th}_x {\rm Be}_{13} $
($ 0.02 \leq x \leq 0.045 $) states appear which probably break
time-reversal symmetry. It was shown theoretically that such superconducting
states can generate spontaneous supercurrents and magnetic field
distributions in the vicinity of lattice defects and surfaces
\cite{IMPURITY}. In
both compounds the occurrence of such local fields in connection
with the superconducting phase transition has been detected by means
of muon spin rotation ($ \mu $SR) measurements
\cite{MSR}. For both compounds,
consistent phenomenological theories for this effect have been
formulated \cite{REVIEW}.

In the field of HTSC, various theories and mechanisms leading to
$ {\cal T} $-violating superconducting states have been proposed.
The effective two-dimensionality of the cuprates may serve as a basis
for particles with fractional statistics, the so-called anyons
\cite{ANYON}. Laughlin
showed that the resulting superconducting state has a composite
order parameter of the form $ d_{x^2 - y^2} + i \epsilon d_{xy} $
which obviously breaks time-reversal symmetry \cite{LAUGH}. Alternative
mechanisms can lead to
a $ {\cal T} $-violating states with the symmetry $ s+id_{x^2 - y^2} $
\cite{SID}. At present there is no indication beyond any
doubt that such states are realized in the HTSC \cite{EXP}.
On the contrary,
recent experiments demonstrate that at least at the onset of
superconductivity ($ T \approx T_c $) the critical behavior of
the London penetration depth is that of a single component order
parameter belonging to the universality class of the XY-spin
model \cite{XY}.  Only below an additional superconducting transition at
lower temperature could the composite $ {\cal T} $-violating order
parameter appear.  No signs of such an additional
phase transition have yet been observed in the thermodynamic properties.
In addition, it should be noted that each of the $ {\cal T} $-
violating states mentioned above lacks gap nodes.
This fact would also
lead to inconsistency with low-temperature measurements of the
London penetration depth which demonstrate the presence
of nodes, as mentioned above \cite{HARDY}.

In this paper we will show that there is no conflict between the
interpretation of the experiment by Kirtley and coworkers
\cite{TRIANGLE} which could
indicate $ {\cal T} $-violation and the other experiments
which obviously rule out the existence of such a state
\cite{SBL}.  We argue that
the latter experiments address bulk properties, while the former one
considers effects in connection with interfaces and grain boundaries.
The seeming conflict is resolved when we assume that
$ {\cal T} $-violation occurs only locally in the immediate vicinity
of an interface. The bulk, on the other hand, may only have a single
component order parameter, presumably with $d$-wave symmetry, but
we cannot rule out other symmetries. As we will discuss below, the
extension of the $ {\cal T} $-violating state towards the bulk is
rather short, of the order of coherence length $ \xi $. 

\section{A first interpretation of the experiment}

Let us now examine the properties of an arrangement
similar to the one used by Kirtley and coworkers \cite{TRIANGLE}.
As illustrated in Fig. 1a, it is a superconducting film of
triangular shape as an inclusion in another
superconducting film, both of the same material.
The crystal symmetry is tetragonal (for simplicity we neglect
here the orthorhombic distortion present in many
HTSC) and the film is $c$-axis textured.  The basal
plane axes (of the inclusion and the surrounding) are misoriented with
each other. The interfaces (the edges of the triangle, each of 
length $ L $) are weak links between the inner and the outer film. For
simplicity we will treat them as Josephson contacts so
that the standard sinusoidal current-phase relation applies.

\subsection{Pure $d$-wave symmetry}

Let us analyze the properties of this arrangement under
the assumption that the superconductor here is a $d$-wave
superconductor with an order parameter symmetry as
the pair wavefunction $ \psi_d({\bf k}) = \cos k_x - \cos k_y $.
This means that
we should carefully consider the intrinsic phase structure of
the order parameter when deriving the Josephson
current-phase relation. The phase difference between the positive
and negative lobes of the pair wavefunction is $ \pi $. If dominant
lobes of the same sign face each other at an interface, the
corresponding Josephson current-phase relation has the standard
form and the interface energy is minimized by a vanishing
difference the order parameter phases (0-junction). However,
if the facing lobes have opposite sign, an additional phase
$ \pi $ enters and the energy is minimized by a phase 
difference of $ \pi $ ($ \pi $-junction) \cite{SRWOHL}.  We have
\begin{equation}
E_J(\varphi) = -\frac{I_c \Phi_0}{2 \pi c}  \cos(\varphi-\alpha)
\end{equation}

\noindent
where $ \alpha = 0 $ for a 0-junction and $ \pi $ for a
$ \pi $-junction, and $ \varphi $ is the phase difference
through the interface. In Fig. 1a we assume that
the edge segments 1-2 and 3-1 can be labeled as a $ \pi $-junction
and segment 2-3 as a 0-junction.  This definition is
not unique. A redefinition of the order parameter phase
in one of the two
superconductors ($ \phi \to \phi + \pi $) would reverse this labeling.

\begin{figure}
\begin{center}
\leavevmode
\epsfbox{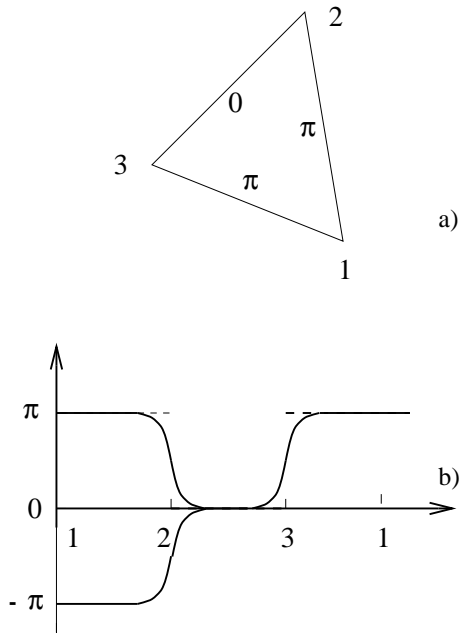}
\end{center}
\vspace{0.1in}
\caption{Triangular grain boundary in a $d$-wave superconductor:
(a) The edges act as Josephson junctions with phase shifts, 0
or $ \pi $; (b) The phase $ \varphi $ tries to be pinned at phase shift
values (indicated by dashed lines) and change in an antikink (kink) of width $
\lambda_J $ at the corners 2 and 3.  There are several
possible solutions for $ \varphi $ due to the $ 2 \pi $-periodicity of
the phase. The solution with an anti-kink and a kink leads 
to an overall
phase winding 0 around the triangle.  An equivalent solution with two
kinks (also shown) would give $2 \pi $ winding.} 
\end{figure}

We now map all segments of the interface onto a one-dimensional
axis with periodic boundary conditions for the coordinate
$ \tilde{x} $ as shown in Fig. 1b ($ \tilde{x} + 3 L = \tilde{x} $).
Here we can study the spatial variation of $ \varphi $ along $ \tilde{x} $
by using the Sine-Gordon equation

\begin{equation}
\partial_{\tilde{x}}^2 \varphi
= \lambda^{-2}_J \sin(\varphi - \alpha(\tilde{x})),
\end{equation}

\noindent
where both the Josephson penetration depth $ \lambda_J =
(\phi_0 c / 8 \pi^2 d I_c)^{1/2} $ ($d$: magnetic width of the
interface)
and the intrinsic phase shift $ \alpha $ are assumed to be
constant within each segment (see Fig.1b)
\cite{TINKHAM}. For good junctions
($ \lambda_J \ll L $), $
\varphi $ tends to be pinned to the value $ \alpha $ in each
segment of the interface, but has to change at the boundaries
where $ \alpha $ is discontinuous. The solution of Eq.(2)
shows kinks at these boundaries with an extension of
$ \lambda_J $ (Fig. 1b). Note that the two kinks at 2 and 3 can
be either ``kink'' and ``anti-kink'' or both ``kinks'' as a consequence
of the periodicity of Eq. (2). Other types of kink solutions
are energetically more expensive.

The spatial variation of $ \varphi $ induces a local magnetic
flux density on the interface given by the expression
$ \phi(\tilde{x}) = \Phi_0 \partial_{\tilde{x}}
\varphi(\tilde{x})/ 2 \pi $
Therefore, each kink corresponds to a local flux line or vortex
with a magnetic flux $ \Phi = \Phi_0 (\varphi_{r} - \varphi_{l})
/ 2 \pi $, where $ \varphi_{r(l)} $ denotes the values of
$ \varphi $ far enough to the right (left) of the kink such that
$ \partial_{\tilde{x}} \varphi $ is essentially zero.
Both kinks in Fig. 1b indicate vortices
with $ \Phi = \pm \Phi_0 /2 $. Due to the periodicity of
$ \tilde{x} $ we find that $ \varphi(\tilde{x} + 3L) =
\varphi(\tilde{x}) + 2 \pi n $ ($ n $: integer), which requires
that the total flux integrated over the whole triangle
interface be an integer multiple of $ \Phi_0 $. 
Of course, the triangle is surrounded
by a superconductor whose single valued order parameter allows
phase windings of $ 2 \pi n $ only. This ``sum rule'' implies
that half-integer flux lines can only appear at two of the
three corners. Because the flux on each corner can only
vary by $ n \Phi_0 $, at corner 2 and 3 there is always a
flux line with a flux of at least $ \Phi_0 /2 $.  Larger
fluxes could be stabilized by an external field.
This result is equivalent to the one presented
in Ref. 8 and 13.

The comparison of our ``experiment'' with the one performed in
reality shows that the simple picture we tried to draw here
does not explain the measurement by Kirtley and coworkers
\cite{TRIANGLE}.
They found fluxes at all corners, all of which are clearly 
smaller than $ \Phi_0 / 2 $. We call them {\it fractional vortices}.
In all samples checked, the sum rule constraining the total 
flux on the boundary to an integer multiple of $\Phi_0$ was 
satisfied with good accuracy.

In the introduction we claimed that the existence of
fractional vortices requires a superconducting phase with
broken time-reversal symmetry. We give here a brief argument
for this statement.  Consider one of the corners of the
triangle (or a similar structure) with a vortex whose flux
is $ \Phi $.  Apply the time-reversal operation to this
system.  This reverses the flux ($ \Phi \to - \Phi $). 
If the superconductor is otherwise invariant
under this operation ($ {\cal T} $-invariant), the difference
between $ \Phi $ and $ - \Phi $ must be an integer multiple
of $ \Phi_0 $ as in every standard superconductor: $ \Phi
= - \Phi + n \Phi_0 $.  Therefore, $ \Phi $ ($ = n
\Phi_0 /2 $) is an integer or half-integer quantum of $ \Phi_0 $
as seen above.  Consequently the observation
of a vortex with a flux different from those values can only mean
that the superconducting state is not invariant under the time-reversal
operation \cite{SYBK}.

\subsection{Josephson effect for a $ {\cal T} $-violating
interface}

A ${\cal T} $-violating superconducting order parameter consists
of at least two components (e.g., $ d_{x^2 - y^2} + i d_{xy} $ 
or $ s + i d_{x^2 - y^2} $).
We therefore restrict ourselves to the case of a two-component order
parameter with a generic pair wave function $ \psi({\bf k}) =
\eta_1 \psi_1 ({\bf k}) + \eta_2 \psi_2({\bf k}) $. Here $ \eta_1 $
and $ \eta_2 $ are the two complex order parameters with the symmetry
properties of the corresponding pair wavefunctions. Time-reversal
transforms the order parameter to its complex conjugate 
($ \eta_j \to \eta^*_j $). If time-reversal symmetry is conserved,
then $ (\eta_1, \eta_2) $ is up to a common phase factor equal to
$ (\eta^*_1 , \eta^*_2 ) $ or $ \eta_1 / \eta_2 = \eta^*_1 / \eta^*_2 $.
Otherwise, the order parameter breaks time-reversal symmetry and the
state is at least two-fold degenerate, since both $ (\eta_1, \eta_2) $
and $ (\eta^*_1 , \eta^*_2 ) $ have the same free energy. We will 
consider here the standard situation where $ \eta_1 / \eta_2 $ is 
imaginary: $ \phi_1 - \phi_2 = \pm \pi/2 $ ($ \eta_j = | \eta_j |
\exp(i \phi_j) $).

For an interface between two superconductors ($ A $ and $ B $)
both with order
parameter $ (\eta^{A(B)}_1 , \eta^{A(B)}_2 ) $ the Josephson current
phase relation has the form

\begin{equation}
J = \sum^2_{i,j=1} J_{cij} \sin(\phi^{B}_i - \phi^{A}_j)
\end{equation}

\noindent
where $ J $ is the supercurrent density at a given point on
the interface. There are four different combinations for the
phase coherent coupling and $ J_{cij} $ denotes the coupling
strength between the components $ \eta^B_i $ on side $ B $ and
with $ \eta^A_j $ on side $ A $ ($ \eta^{\mu}_j =
|\eta^{\mu}_j| \exp(i \phi^{\mu}_j) $) with $ j=1,2 $ and $ \mu
=A,B $). The energy for a uniform interface of area $ S $ is
given by

\begin{equation}
E_J = - \frac{\Phi_0 S}{2 \pi c} \sum^2_{i,j=1} J_{cij}
\cos(\phi^{B}_i - \phi^{A}_j)
\end{equation}

\noindent
For simplicity we restrict ourselves to the situation where
$ \phi^A_1 - \phi^A_2 = \phi^B_1 - \phi^B_2 = \pi/2 $ is fixed on
both sides of the interface. (This
phase difference can be different in the vicinity of the junction without
changing the conclusion we will draw here for the simplified case.)
The current density $ J $ and the interface energy density
$ \epsilon_J $ depend
only on one phase difference through the interface, say 
$ \varphi = \phi^B_1 - \phi^A_1 $.

\begin{equation} \begin{array}{l}
\displaystyle
J(\varphi) = \tilde{J}_c \sin(\varphi - \tilde{\alpha}) \\ \\
\displaystyle
\epsilon_J (\varphi) = -\frac{\Phi_0 \tilde{J}_c S}{2 \pi c}
\cos(\varphi - \tilde{\alpha}) \\
\end{array} \end{equation}

\noindent
with

\begin{equation} \begin{array}{l}
\displaystyle
\tilde{J}_c = \sqrt{(J_{c11} + J_{c22})^2 + (J_{c12} - J_{c21})^2 } \\ \\
\displaystyle
\tan (\tilde{\alpha}) = \frac{J_{c12} - J_{c21}}{J_{c11} + J_{c22}} \\
\end{array} \end{equation}

\noindent
The phase shift $ \tilde{\alpha} $ corresponds to $ \varphi $
minimizing the interface energy. In this sense the correct solution
for $ \tilde{\alpha} $ must be chosen in Eq. (6) \cite{GESH}.
Obviously,
$ \tilde{\alpha} $ can assume any value and depends only
on the relative magnitude of the different coupling components
which parameterize here the interface properties. It is easy to
follow the same consideration for the $ {\cal T} $-invariant
combination of $ \eta_1 $ and $ \eta_2 $ where we find that
$ \tilde{\alpha} $ is strictly either 0 or $ \pi $.

Let us apply this result to the triangle studied above, assuming
$ {\cal T} $-violating superconducting states. Each of the interface
segments as a uniform junction is characterized by a phase shift
$ \tilde{\alpha} $ ($ 0 < \tilde{\alpha} < \pi $). Analyzing Eq. (2)
for this situation we find that there are now kinks
of $ \varphi $ at all three corners connecting the different values
of $ \tilde{\alpha} $ in each segment (Fig.2). This leads to vortices at
these corners whose fluxes are given by

\begin{equation}
\Phi = \Phi_0 \frac{\tilde{\alpha}_r - \tilde{\alpha}_l}{2 \pi}
\end{equation}

\noindent
with $ \tilde{\alpha}_{r(l)} $ as the values of $ \tilde{\alpha} $
on the segments on right (left) of the corner. It is obvious that
in this case the flux $ \Phi $ need not be $ \Phi_0 $ or
$ \Phi_0/2 $. Thus, these kinks correspond to what we described as
fractional vortices above. Note, that also here the sum of all
fluxes must add up to an integer multiple of $ \Phi_0 $ because of the
periodic boundary condtions for $ \varphi $.

\begin{figure}
\epsfbox{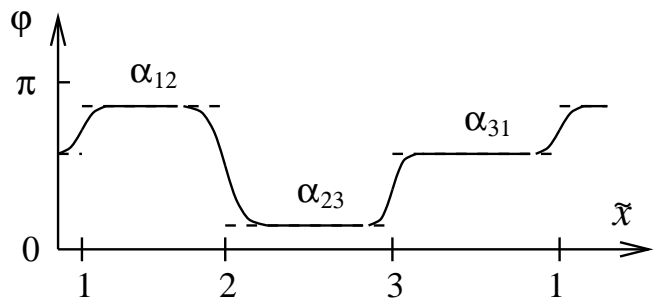}
\vspace{0.1in}
\caption{Phase $ \varphi $ for a triangle with $ {\cal T} $-violating
  interface states. The phase shifts $ \alpha_{ij} $ are indicated by
  the dashed line. $ \varphi $ follows these phase shifts by creating
  (anti-)kinks at all three corners. The kink heights are different
  from a multiple of $ \pi $ in general and lead to fractional fluxes.}
\end{figure}
\section{The interface state}

As mentioned earlier, the assumption of a $ {\cal T} $-violating bulk
superconducting state is incompatible with a number of experiments.
It is, however, possible that $ {\cal T} $-violation occurs only in
the vicinity of the interface (grain boundary).  This is enough to 
generate fractional vortices at grain boundary corners.  In this
section we would like 
to discuss such an interface state based on a Ginzburg-Landau theory.

\subsection{Ginzburg-Landau theory}

We use a Ginzburg-Landau free energy functional of two complex
order parameters $\eta_1$ and $\eta_2$.  The first order parameter
shall be the dominant one corresponding to the $d_{x^2-y^2}$-wave pairing
state. For the second we choose the $d_{xy}$-wave pairing state. 
Thus the pairing state is a combination of the two components

\begin{equation}
\psi({\bf k}) = \sum_{i=1,2} \eta_i \psi_i ({\bf k})= \eta_1 \psi_1
({\bf k}) + \eta_2 \psi_2 ({\bf k})
\end{equation}
with

\begin{equation}
\psi_1 ({\bf k}) = k^2_x - k^2_y \qquad \mbox{and} \qquad 
\psi_2 ({\bf k}) = k_x k_y.
\end{equation}

The free energy functional must be invariant under the symmetry
transformations of the crystal group (in this case tetragonal,$D_{4h}$),
time reversal and U(1)-gauge symmetry.  
Keeping terms up to fourth order in the expansion with respect to
order parameter, we have

\begin{equation}
{\cal F} = \int d^2x\, [ {\cal F}_I + {\cal F}_{12} + {\textstyle
\sum_{i=1,2}} {\cal F}_i ]
\end{equation}

\noindent
where

\begin{equation}
{\cal F}_i  =  \alpha_i
|\eta_i|^2 + \frac{\beta_i}{2} |\eta_i|^4 + K_i |{\bf D}\eta_i|^2
\end{equation}
\begin{eqnarray}
{\cal F}_{12} &  = &
\gamma|\eta_1|^2|\eta_2|^2 + \delta(\eta_1^2 \eta_2^{*2} + \, \mbox{c.c.})
+ \kappa^2 (\nabla \times {\bf A})^2 \nonumber \\
& + & \tilde{K} \left( (D_x \eta_1)^*(D_y \eta_2) - (D_y \eta_1)^*
(D_x \eta_2) + \,\mbox{c.c.} \right)
\end{eqnarray}
\begin{eqnarray}
{\cal F}_I  & = & \sum_{i=1,2} \{ g_i^{A,B} |\eta_i^{A,B}|^2 
\nonumber \\
& - & \frac{\Phi_0}{2 \pi c} 
\sum_{j=1,2} \tilde{J}_{ij} {\rm Re} \, \eta_i^B \eta_j^{*A} \}
       \delta({\sl interface}).
\end{eqnarray}

We work in units where ${\bf D}=\nabla-i 2 \pi {\bf A}/\Phi_0 $ ($
{\bf A} $: vector potential).  The coefficients are
all phenomenological parameters which contain the relevant information of
microscopic origin. The second order coefficients depend on the
temperature in the usual way ($ \alpha_i = a'(T-T_{ci}) $ where $ T_{ci} $
is the bare transition temperature of $ \eta_i $).  The interface free
energy ${\cal F}_I$ includes the pair breaking effect in the
first term and the Josephson junction energy in the second term (note:
$ \tilde{J}_{12} = \tilde{J}_{21} $). The superscripts $B$ and $A$ denote the
different sides of the interface, or ``outside'' and ``inside'' if the
interface is a closed curve. 

It is worth noting that the free energy expansion without interface term is
cylindrical symmetric with $ z $ as the rotation axis, although the system has
only tetragonal symmetry.  Anisotropy enters via the interface
term. The coefficients $ g_i $ and $ \tilde{J}_{ij} $ depend in
general on the angle between crystal axes and the interface. Pair
breaking for the $ d_{x^2 - y^2} $-wave ($ d_{xy} $-wave) phase is
most effective when 
the interface normal vector points along the [1,1] ([1,0])-direction,
i.e. along the nodes of the pair wavefunction \cite{REVIEW}. On the
other hand, pair 
breaking is weaker when the lobes of the pair wavefunctions point
towards the interface. Consequently, the $ d_{x^2-y^2} $-wave
component of the order parameter is suppressed weakly where the $
d_{xy} $-wave component is affected most (and vice versa). Similarly, the
interface coupling coefficients depend on the internal structure of
the pair wavefunction. If we denote the angle between the interface
normal vector and the crystalline $ x $-axis by $ \vartheta $, they are
$ \tilde{J}_{ij} = J_0 f_i(\vartheta_B) f_j(\vartheta_A) $, where the
functions have the generic angular structure
$ f_1 (\vartheta) = \cos (2 \vartheta) $ and $ f_2 (\vartheta)
= \sin (2 \vartheta) $ which indicate the internal 
phase structure of the two pair wavefunctions.

The variation of $ {\cal F} $ with respect to $ \eta_i $ and $ {\bf A}
$ yield the following Ginzburg-Landau equations:

\begin{eqnarray}
K_i {\bf D}^2 \eta_i & = & \alpha_i \eta_i
+ \beta_i |\eta_i|^2 \eta_i + \gamma |\eta_{3-i}|^2
\eta_i \nonumber \\
& + & 2\delta \eta_{3-i}^2 \eta_i^* - (-1)^i \tilde{K} B_z \eta_{3-i}
\end{eqnarray}
and

\begin{eqnarray}
\kappa^2 \bbox{\nabla} \times ( \bbox{\nabla} \times {\bf A})= 
\kappa^2 {\bf j} & = & \sum_{i=1,2} K_i ({\rm Im} \, \eta_i^* \nabla \eta_i -
|\eta_i|^2 {\bf A} ) \nonumber \\
& + & \tilde{K} \nabla \times (\hat{z}\,{\rm Im} \, \eta_1 \eta_2^*)
\end{eqnarray}
The term $ {\cal F}_I $ generates the boundary condition at the
interface with normal vectors $ {\bf n}^{A,B} $,

\begin{eqnarray}
K_i {\bf n}^A \cdot (\bbox{\nabla} - i \frac{2 \pi}{\Phi_)}{\bf A})
\eta^A_i & = & 
g^A_i \eta^A_i - \frac{\Phi_0}{2 \pi c}\sum_{j=1,2} J_{ij} \eta^B_j
\nonumber \\ 
K_i {\bf n}^B \cdot (\bbox{\nabla} - i  \frac{2 \pi}{\Phi_)} {\bf A})
\eta^B_i & = & 
g^B_i \eta^B_i - \frac{\Phi_0}{2 \pi c} \sum_{j=1,2} J_{ij} \eta^A_j
\end{eqnarray}
The imaginary part is related to the expression for the supercurrent $ j $
perpendicular to the interface and leads to the Josephson current
(Eq. (3)) with

\begin{equation}
J_{cij} = J_{ij} | \eta^B_i | | \eta^A_j |
\end{equation}
where the order parameter values are taken at the interface. 

Before discussing the interface problem, we consider the properties of
the order parameter in the bulk. We assume that $ T_{c1} > T_{c2} $
such that for temperatures immediately below $ T_{c1} $ only the
component $ \eta_1 $ becomes finite, while $ \eta_2 $ remains zero. 
The instability condition for the occurrence of $ \eta_2 $ is given by
 
\begin{equation}
\alpha_2(T^*) + ( \gamma - 2 | \delta |) | \eta_1 (T^*) |^2 = 0,
\end{equation}
defining the transition temperature $ T^* < T_{c1} $. This second
transition leads to a state where both $ \eta_1 $ and $ \eta_2 $ are
finite. The relative phase, $ \theta = \phi_1 - \phi_2 $, depends on
the sign of $ \delta 
$. For $ \delta < 0 $ the combination is real, $ \theta = 0 $ or $ \pi
$ and for $ \delta > 0 $ $ \theta = \pm \pi/2 $. The latter state
breaks time reversal symmetry. For the following we will assume that 
$ \delta > 0 $. However, the other parameters shall be chosen so that
$ T^* \leq 0 $ in order to avoid the second transition as it is not
observed in the experiment. We also require that $ \gamma > 2 \delta $
so that the two order parameter components tend to suppress each
other.

\subsection{Order parameter}

We consider now an infinitely extended interface. In this case the
order parameter and the vector potential only depend on the coordinate
perpendicular to the interface. With certain simplifications a
qualitative discussion is possible as we showed in Ref.\cite{SBL}. We
would like to present here an analytical study and then substantiate 
the result by a complete numerical treatment.

The basic concept leading to an unconventional
interface state is the following. Pair
breaking at the interface reduces the $ \eta_1 $-component locally. It
recovers, however, over a coherence length $ \xi = \sqrt{K_1 /
  |\alpha_1|} $. It is easy to see from Eq. (18) that a local reduction
of $ \eta_1 $ leads to a local enhancement of $ T^* $ (note: $ \gamma
> 2 \delta $). Consequently,
the $ \eta_2 $-component can appear at the interface at sufficiently
low temperature ($ T < T' $), but it decays exponentially towards the bulk. The
extension of $ \eta_2 $ diverges when $ T $ approaches the bulk
$ T^* $. Therefore $ \eta_2 $ does not possess the same length
scale as $ \eta_1 $ in general. Because $ \delta > 0 $, the combination
of $ \eta_1 $ and $ \eta_2 $ is complex $ \theta \neq 0 $ or $ \pi $.

Let us now consider a simplified analytic solution of the Ginzburg-Landau
equations for an interface whose normal vector is parallel to the $x$-axis.
We neglect the vector potential and the coupling
between the two sides ($ \tilde{J}_{ij} = 0 $). A qualitatively good view of
the interface state is obtained in the limit where the length scales
$ \xi_i $ of the order parameters are very different, i.e. for
$ \xi_1 \ll \xi_2 $. At the interface, $ \eta_1 $ behaves approximately
like

\begin{equation}
  \eta_1 (x) = \eta_{10} {\rm tanh} \left(\frac{|x|+x_0}{\xi_1}
  \right)
\end{equation}
as can be found by solving the Ginzburg-Landau equation with $ \eta_2
= 0 $. The boundary condition at the interface determines $ x_0 $ and
$ \eta_{10} = \sqrt{|\alpha^*|/2 \beta_1} $ is the bulk value of the
order parameter. Fixing $ \eta_1(x) $ the equation for $ \eta_2 $
becomes

\begin{eqnarray}
K_2 \partial^2_x \eta_2  & = & (\alpha_2 + \gamma 
\eta^2_{1}(x)) \eta_2 +2 \delta \eta^2_{1}(x) \eta^*_2 \nonumber \\
& + & \beta_2 |\eta_2|^2 \eta_2 + g_2 \eta_2 \delta (x) 
\end{eqnarray}
Following our assumption about the coherence lengths, we approximate 
the spatial dependence of $ \eta_1 (x) $ by a $ \delta $-function
in this equation. It is easy to see that $ \eta_2 $ is purely
imaginary, $ \eta_2 = i u(x) $ and satisfies the equation

\begin{equation}
K_2 \partial^2_x u = \alpha^* u + \beta_2 u^3 + (g_2 - (\gamma - 2
\delta) \eta^2_{10} s \xi_1 ) \delta(x)
\end{equation}
The factor $ s $ is of the order one. The solution is a hyperbolic function,

\begin{equation}
u(x) = \sqrt{\frac{2 \alpha^*}{\beta_2}} \frac{1}{{\rm
    sinh}((|x|+\tilde{x}_0)/\xi_2)}
\end{equation}
where $ \xi^2_2 = K_2 / \alpha^* $. We use the boundary condition at the
interface to find the shift $ \tilde{x}_0 $ ($ > 0 $),

\begin{equation}
{\rm coth} \left(\frac{\tilde{x}_0}{\xi_2} \right) = 
\frac{(\gamma - 2 \delta) \eta^2_{10} \xi^2_2 s - g_2 \xi_2}{2 K_2}.
\end{equation}
Setting the right hand side equal one determines the critical
temperature $ T' $ for the occurrence of $ \eta_2 $. At this
point $ \tilde{x}_0 = \infty $. Note that $ T' > T^* $ and that
$ \eta_2 $ extends into the whole superconductor when we approach
$ T = T^* $. In the range $ T^* < T < T' $ it decays exponentially
on the length $ \xi_2 $.
Our numerical results below will show that this approximate solution
describes the interface state well apart from the fact that we do
not resolve here variations on length scales of $ \xi_1 $.

For finite $ \tilde{J}_{ij} $ we can determine the phase shift $\alpha $. 
In this symmetric formulation of the interface problem the phase shift
$ \alpha $ would be strictly 0 or $ \pi $ even if the state is a
complex combination of $ \eta_1 $ and $ \eta_2 $, i.e. it breaks time
reversal symmetry. We find
immediately that Eqs. (6) and (17) lead to $ \alpha = 0 $ or $ \pi $,
because $ \tilde{J}_{12} = \tilde{J}_{21} $. 
A condition sufficient for $ \alpha $ different from these
trivial values would be different coefficients $ g_i $ on both sides
of the interface, which means for example that the crystal orientation
on side A and B is different (as it is for a grain boundary).  It can 
be shown that the violation of ``parity'', the mirror symmetry due to 
reflection at the interface, is necessary (see \cite{SYBK}).

\subsection{Magnetic properties} 

The interface state is an inhomogeneous $ {\cal T} $-violating
superconducting state for temperatures below $ T' $. It shows unusual
magnetic features which 
originate from an orbital magnetic degree of freedom of the Cooper
pairs. Remember that the two pairing components we consider both belong to
the $d$-wave channel. If they are combined with a relative phase
different from $ \theta = 0,\pi $, then the pairing state has a
component belonging to the spherical harmonic $ Y_{2,\pm 2} ( 
{\hat{\bf k}}) \propto (k_x \pm i k_y) $. This Cooper pair state has a
  finite orbital magnetic moment $ M \hat{{\bf z}} $ 
parallel to the $ z $-axis. The moment generates circular currents in
the $ x $-$ y $-plane which cancel in the uniform
superconducting phase. However, in the inhomogeneous region at the interface
they can appear as finite currents, $ {\bf j} = \bbox{\nabla} \times
\hat{{\bf z}} M ({\bf r}) $. These currents are included within our
phenomenological description. 

Let us discuss the magnetic part of the Ginzburg-Landau equation,
Eq. (14). Due to the translational invariance parallel to the interface,
no currents are allowed to flow through the interface in the
energetically lowest state (otherwise, the Josephson energy would
not be minimized). Through the equation for the $ A_x $-component, 
we find that this requires $ A_x = 0 $. The $ A_y $-component then 
satisfies the following equation,

\begin{equation}
\partial^2_x A_y - \lambda^{-2} A_y = -\frac{\tilde{K}
  }{\kappa^2} \partial_x  \eta_1 \eta_2 
\end{equation}
where $ \lambda^{-2} = K_2 \eta^2_{10} / \kappa^2 $ is the London
penetration depth. The right hand side denotes the current due to the
magnetic moment, which is proportional to $ i \hat{{\bf z}} ( \eta^*_1
\eta_2 - \eta_1 \eta^*_2) $. This current $ j_y $ flows parallel to the
interface. We do not discuss the influence of the vector potential on
the order parameter here.
Inserting the solution found above for the interface state, we find

\begin{equation} \begin{array}{ll}
j_y = & \displaystyle 
- \frac{\tilde{K} \eta_{10}}{\kappa^2}  \sqrt{\frac{2
    \alpha^*}{\beta_2}} \left[ 
\frac{{\rm sinh}((|x|+ \tilde{x}_0)/\xi_2) }{
  {\rm cosh}^2 ((|x|+ \tilde{x}_0)/\xi_2)} \frac{|x|}{x \xi_2} \right. \\ & \\
& \left. \displaystyle - \frac{s \partial_x \delta (x)}{{\rm sinh}(
  \tilde{x}_0 / \xi_2 )} \right] . \\
\end{array} \end{equation}
This current distribution is odd under reflection through the interface.
Starting at zero on the interface, it rises quickly and has a maximium
at a distance of about $ \xi_1 $. Then it changes sign and decays on
the length scale $ \xi_2 $. The magnetic field generated by this
current has a narrow peak of width $ \xi_1 $ on the interface followed
by two wings of opposite sign. Within our approach we obtain

\begin{equation}
B_z =  \frac{\tilde{K} \eta_{10}}{\kappa^2} (u_2(x) - s \delta(x) )
\end{equation}
where we neglect the screening effects due to the second term on the
left hand side of Eq.(24). Because the field in Eq.(26) leads to a
finite magnetic flux, screening currents are induced which yield a
compensating diamagnetic field with the length scale $ \lambda $. 
This screening effect can be rather small if the two contributions 
in Eq.(26) nearly cancel each other. The net magnetization vanishes
exactly, because in the interior of a superconductor phase coherence
allows for a net magnetic flux only if there is a winding of the order
parameter phase. 

\begin{figure}
\epsfbox{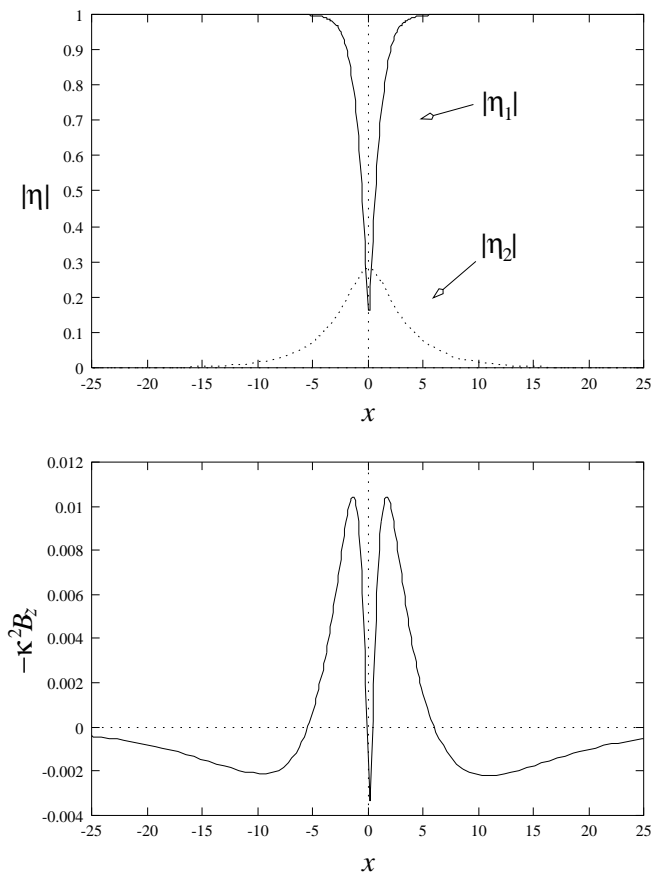}
\vspace{0.1in}
\caption{Numerical solution of the infinite interface state.  {\sl Top:}
The modulus of the two order parameter components. The relative 
phase is constant $\pi/2 $.  {\sl Bottom:}  
The $ z $-component of the magnetic 
field (multiplied by $\kappa^2$).  In these units, $\xi_1=1$ and 
$\lambda=7$ ($\kappa^2=49$).  The magnetic field exhibits variations 
on the lengthscales $\xi_1$ and $\xi_2$ in the immediate vicinity 
of the interface, 
and is screened away from the interface on the lengthscale $\lambda$. }
\end{figure}

\subsection{Numerical solution for the infinite interface state}

Finally we solve the complete set of coupled GL equations for the
uniform interface in order to show that our analytic treatment gives
the correct qualitative behavior. We choose the coherence length $
\xi_1 = \sqrt{K_1 / | a_1|} $ as the unit length and the London
penetration depth about 10 times $ \xi_1 $. For the coefficients we
use values of order unity: $ a'= \beta_2 = \beta_1 = K_2 = K_1
=1, \tilde{K}=0.1, \gamma=1 $ and $ \delta = 1/4 $. The transition
temperature of the $ d_{xy} $-component is $ T_{c2} = 0.4 T_{c1} $. 
For the space coordinate we introduce a fine mesh and the interface is
taken as single point. For simplicity we consider the symmetric
situation by representing the interface via a local suppression of $
T_{c1} $: $ g_1 |\eta_1|^2 \delta(x) $.  

The solution for $g_1 =4, g_2=0$ is shown in Fig. 3. 
The shapes of the order parameters
(top) are in good qualitative agreement with the analytic result. 
We see that the length scales $ \xi_1 $ and $ \xi_2 $ are different
and that the latter is larger than the former. The
relative phase $ \theta $ between the two components is constant,
$ \theta = \pi/2 $. The magnetic field has a narrow peak with a width
of the order $ \xi_1 $ on the interface (bottom). Towards the bulk, 
the field changes sign and decays on the length $ \lambda $ 
($ = 7 \xi_1 $). The positive and negative parts of the field 
distribution cancel each other such that no net magnetization is 
present, as anticipated above. 

\subsection{Numerical solution for the grain boundary corner}

In Sect. II we showed that at the border between two different
grain boundary segments with $ {\cal T} $-violating states a vortex
with fractional flux can appear. We would like to demonstrate this
fact here by solving the complete GL problem for a system with a 
grain boundary which has 
a corner. For simplicity we use a right angle corner which matches
well with our choice for the 2D-square lattice mesh. A variable grid
has been introduced to enhance the accuracy in the vicinity of the
interface where the order parameter and magnetic field have the
largest variations. A steepest descent (relaxation) method was used to
minimize the GL free energy with open boundary conditions at the
borders of the mesh.

The coefficients of the GL free energy are the same in both
grains (see caption of Fig. 5). However, the interface terms for the
grain boundary 
are different for the two segments separated by the corner.
The geometry of the grain boundary configuration is shown in 
Fig. 4.  The $A$ and $B$ superconducting regions are separated
by the grain boundary, represented as a thick black line.
For the segment of the interface 
parallel to the $ x_A $- ($ y_A $-)axis we choose
$ \tilde{J}_{11} = 0.1 $, $ \tilde{J}_{12} = -(+)0.2 $, 
$ \tilde{J}_{21} = -(+)0.2 $ and $ \tilde{J}_{22} = 0.1 $.
In region $A$, $g_1=2.5$ and $g_2=0$.  In region $B$, $g_1=5$
and $g_2=0$.  As discussed above, the difference in these boundary 
conditions yields different phase shifts $ \alpha $ in the two segments.
For the parameters chosen, the phase shifts are equal in magnitude and
opposite in sign in the two segments, as expected from Eqs. 6 and 7.

\begin{figure}
\begin{center}
\leavevmode
\epsfxsize=7.5cm \epsfbox{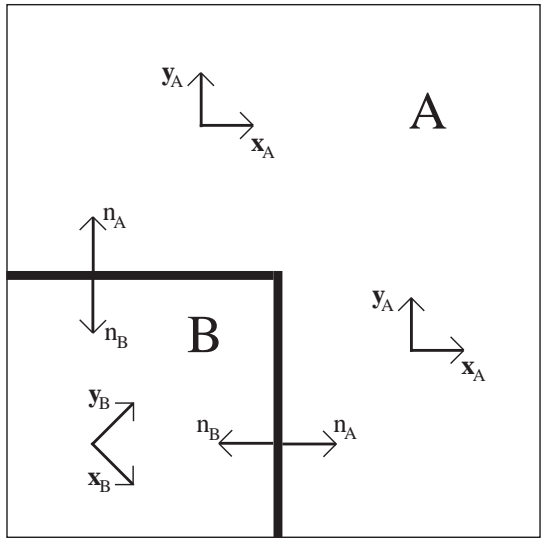}
\end{center}
\caption{Grain boundary configuration used in the calculation of
fractional flux.  The grain boundary
is indicated by the thick line.  The $A$ region 
and $B$ region represent different crystal grains, characterized
by their respective principal axis vectors as shown.  The normal
vectors at each interface are also shown.  In the numerical 
calculation, the density of mesh points is greatest near the 
grain boundaries.}
\end{figure}

In Fig. 5 we show the magnetic field distribution around the corner
which has a pronounced peak indicating the position of the vortex. 
The flux is fractional of the size $ \Phi \approx -0.15 \, \Phi_0 $. 
Along the grain boundary the small field peak occurs which we found
already for the uniform boundary case.  This peak is distorted somewhat
by the Josephson coupling of the order parameters across the interface.
We also see a slight difference
in the decay of the magnetic field towards the grains and along the
grain boundaries. The grain boundaries are usually good contacts so
that the magnetic flux at the corner is well defined.  The sign and
magnitude of the flux can be manipulated by tuning $g_i$ and $J_{ij}$.
Setting all the $J_{ij}$ to 0 eliminates the flux, of course.  Switching
the signs of $J_{12}$ and $J_{21}$ while leaving $J_{11}$ and $J_{22}$
unchanged reverses the sign of the magnetic flux, as expected from Eqs.
6 and 7.  Interestingly, by choosing the $J_{ij}$ and $g_i$ appropriately
we can favor the formation of a domain wall along the grain boundary
at one or both interfaces.  The magnetic field distribution along the
grain boundary in this case becomes antisymmetric with respect to 
reflection across the boundary.  The parameters chosen for Fig. 5
do not lead to such a domain wall on either interface.

The grain boundaries themselves are naturally good conduits for 
magnetic flux.  In the case shown in Fig. 5, negative magnetic flux
pours into the central dip in $B_z$ along the interface and drags
down the side lobes (Fig. 3, bottom).  In general, we expect
the grain boundary to pin bulk vortices ($\Phi = \pm n\Phi_0$) on top
of the fractional vortices due to this tendency to grab magnetic
flux.  This is consistent with the experimental observations of
Kirtley and coworkers \cite{TRIANGLE}.  

\begin{figure}
\epsfbox{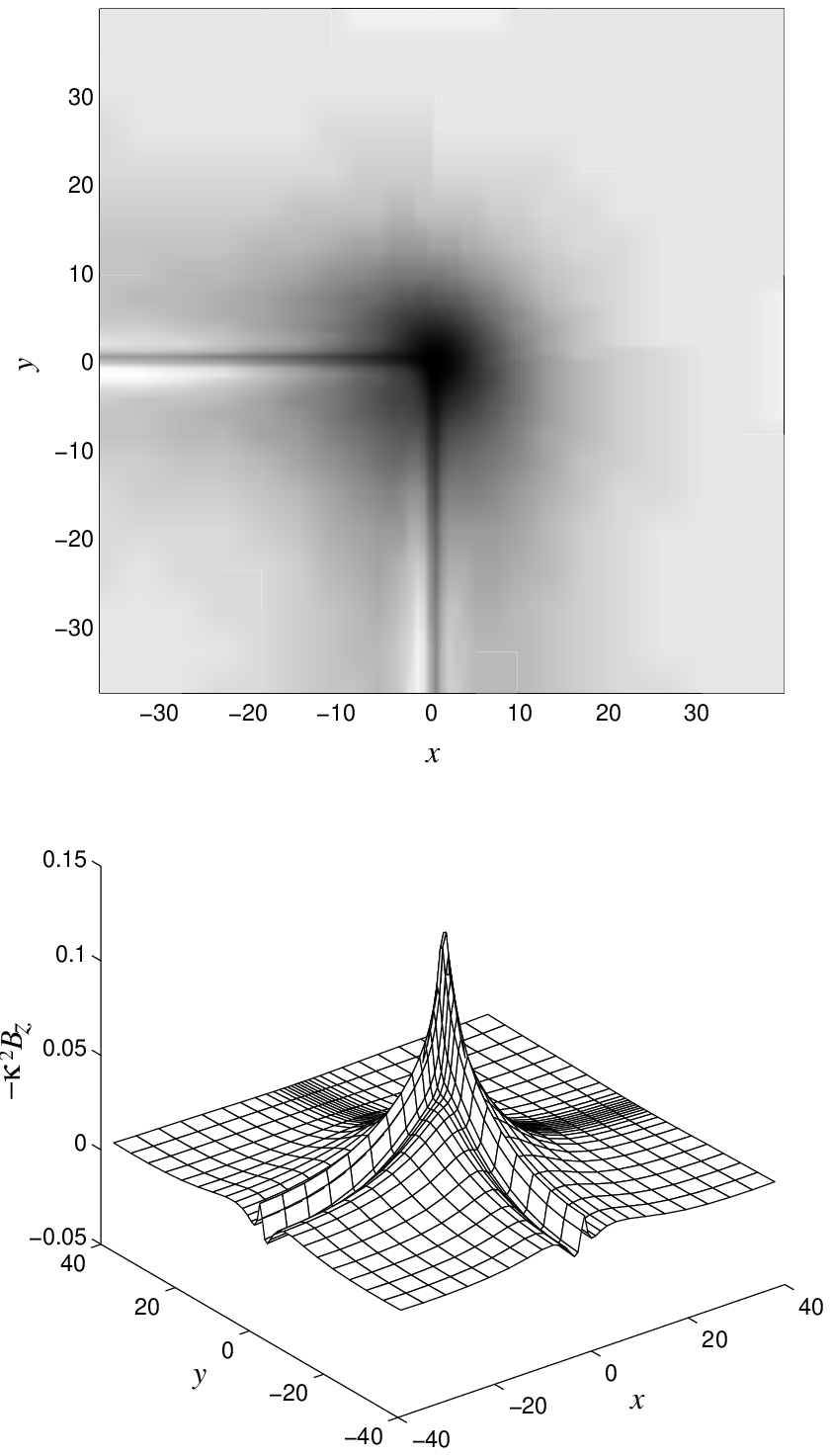}
\vspace{0.1in}
\caption{The magnetic field distribution at this grain 
boundary corner.  {\it Top}: contour plot.  Lighter (darker) 
shades indicate more positive (negative) magnetic fields.
{\it Bottom}: surface plot of $-\kappa^2 B_z$.
The parameters used
are $\xi_1=1$, $\kappa=7$, $K_1=K_2=\alpha_1=\beta_1=\beta_2=\gamma=1$, 
$\delta=1/4$, $\tilde{K}
=0.1$, and $\kappa=7$.  For the segment of the interface 
parallel to the $ x_A $- ($ y_A $-)axis,
$ \tilde{J}_{11} = \tilde{J}_{22} = 0.1 $, $ \tilde{J}_{12} =
\tilde{J}_{12}= -(+)0.2 $. 
In region $A$, $g_1=2.5$ and $g_2=0$.  
In region $B$, $g_1=5$ and $g_2=0$. The enclosed flux is $\Phi 
\approx -0.15 \, \Phi_0$.}
\end{figure}

\section{Conclusions}

We have studied the properties of grain boundaries in $d$-wave
superconductors.  We have demonstrated how they can support a 
superconducting state with broken
time reversal symmetry. We established a connection between such a
state and the existence of vortices carrying a fractional flux. The
key to this connection lies in the observation that a $ {\cal T}
$-violating state at the grain boundary can lead to a non-trivial
phase shift in the Josephson current-phase relation. Vortices occur at
locations where this phase shift changes. Corners of grain boundaries
are important places for such vortices, because they separate segments
with different properties. 

Our results compare qualitatively well with the experimental
observation by Kirtley and coworkers \cite{TRIANGLE}. At present it
is, however, unclear whether we should consider these experimental results
as an evidence for fractional vortices and, consequently, for the
presence of a $ {\cal T} $-violating superconducting phase. Obviously,
if the length scale of the magnetic field along the grain boundary is
comparable with the distance between the corners or the vortices, then
it is impossible to associate a definite flux with each vortex
separately. In order to draw a firm conclusion, this point has to be
clarified experimentally. 

The conditions for the observation of this effect are best if the
crystal orientations of the grains are chosen so that the grain
boundary faces a lobe of the $ d_{x^2-y^2} $-wave pair wavefunction on
one side and nearly a node on the other. As we pointed out in
Sect. III the latter boundary provides a particularly good situation
for pair breaking of the $ d_{x^2-y^2} $-wave component, which is an
important condition for our scenario. Similar conclusions where found
with alternative mechanisms \cite{YIP,KS}. 
Unfortunately, grain boundaries of this type are intrinsically
difficult to produce as homogeneous interfaces and are often hampered
by irregularities. 

In conclusion we would like to emphasize that the effect discussed
here is a consequence of the exotic nature of the superconducting
order parameter. No analoguous effect is possible in the case of a
conventional $s$-wave superconductor, because grain
boundaries have little effect on this pairing type.

We benefited from discussions with P.A. Lee, Y.B. Kim,
K. Kuboki, S. Bahcall and S. Yip.  This work was
supported in part by NSF Grant No. DMR-9120361-002 and the NSF MRL
program through the Center for Materials Research at Stanford Univerity.
M. S. gratefully acknowledges a fellowship from Swiss Nationalfonds and
financial support from the NSF-MRSEC grant DMR-94-00334.


\begin{references}
\bibitem{ABRIKOSOV}
An alternative scenario was recently introduced by A. Abrikosov,
where the occurrence of the node structure in the pair wavefunction
is intimately connected with the orthorhombic distortion;
A.A. Abrikosov, Physica C {\bf 222}, 191 (1994); preprint.
\end{references}
\end{document}